\documentclass[aps,pra,twocolumn,groupedaddress,showpacs]{revtex4}
\usepackage{graphicx}
\usepackage[dvipsnames,usenames]{color}
\usepackage{color}
\usepackage{amsmath}

\begin{document}

\title{Photon bursts at lasing onset and modeling issues in micro-VCSELs}

\author{T. Wang}
\affiliation{School of Electronics and Information, Hangzhou Dianzi University, Hangzhou, 310018, China}
\author{G.P. Puccioni}
\affiliation{Istituto dei Sistemi Complessi, CNR, Firenze, Italy}
\author{G.L. Lippi}
\affiliation{Universit\'e de la C\^ote d'Azur, Institut de Physique de Nice, UMR 7010 CNRS, France}
\email{gian-luca.lippi@inphyni.cnrs.fr}

\begin{abstract}
Spontaneous photon bursts are observed in the output collected from a mesoscale semiconductor-based laser near the lasing threshold.  Their appearence is compared to predictions obtained from Laser Rate Equations and from a Stochastic Laser Simulator.  While the latter is capable of predicting the observed large photon bursts, the photon numbers computed by the former produces a noisy trace well below the experimentally detectable limit.  We explain the discrepancy between the two approaches on the basis of an incorrect accounting of the onset of stimulated emission by the Rate Equations, which instead are capable of complementing the physical description through topological considerations.
\end{abstract}

\maketitle

\section{Introduction}

The steady-state nanolaser response, i.e., the S-shaped emission curve (in double-logarithmic scale) as
a function of energy supplied, is a general feature which has been predicted a
long time ago and which is consistently observed on all devices.  The
averaging process, on which this characteristic curve rests, hides, however, the
details of the interaction and poses questions for which definite answers
have remained elusive for nearly 40 years.  These questions are mainly fundamental, but carry ponderous applicative weight, since the smooth character of the transition between the incoherent and the coherent emission regimes -- all the more so as the the system approaches {\it thresholdless} operation (i.e., very small cavity volume) -- prevents a
clear identification of the {\it laser threshold}:  the pump value at which
the emission is characterized by fully Poissonian statistics.  The latter is a
crucial feature not only in a proper description of lasing, but also for the
practical use of a device, since photon coherence is required to obtain proper interference.  Since most optically-based logical operations are nowadays conducted interferometrically (microrings, Mach-Zehnder interferometers, etc.), the
potential use of nanolasers as light sources for optical interconnects and
all-optical circuits~\cite{Smit2012,Mayer2016,Soref2018,Norman2018,Pan2018}
heavily rests on their coherence properties:  a nanolaser emitting only
partially coherent light would produce poor contrast in the logical operations
and therefore noisy results.  New applications are emerging also in the field of quantum information, where nanolasers with suitable coherence properties may become an integrated element for the generation of single photons on demand~\cite{Kreinberg2018}.  As in the previous examples, mastering the emission properties of the source becomes of paramount importance to ensure system's reliability. 

While it is on one hand possible to obtain (fully,  or nearly fully) coherent emission by suitably increasing the pump -- at least from nanolasers which can withstand strong pump rates --, on the other hand two
difficulties accompany this choice:  1.  the larger the pump, the larger the amount of
energy needed (thus larger costs), and the larger the amount of deposited heat -- a feature
which is definitely adversarial to large-scale component integration --; 2.
the device lifetime may be affected by the stress imposed by the large pump (current)
and by the ensuing thermal load.  

At the present time, direct measurements on
nanolasers are limited to photon counting -- thus to statistical information --, with
the exception of signals acquired by Transition Edge Detectors which can provide a response
linearly proportional to the impinging photon flux~\cite{Schlottman2018} (but
with bandwidth insufficient to resolve the temporal dynamics).  Thus, the
interpretation of the coincidence measurements, translated into correlation
functions, requires a good theoretical model describing laser operation in the
threshold region.  In spite of considerable efforts, and remarkable successes,
in obtaining theoretical descriptions which gain in sophistication either through
detailed microscopic
modeling~\cite{Gies2007,RoyChoudhury2009,Lorke2013,Chow2014,Leymann2015,Jahnke2016,Gies2017},
or through the capture of the relevant physical features with phenomenological
approaches~\cite{Moelbjerg2013,Lingnau2015,Lettau2018,Mork2018} -- including hybrid
microscopic-dynamical descriptions~\cite{Redlich2016,Protsenko2017} --, a
complete representation of the interactions of photons with the gain medium at threshold is
still missing.  One particular feature, which we think holds an important key
for understanding the physics of lasing onset, is the transition from
spontaneous into stimulated emission.  Experimental
measurements~\cite{Wang2015} conducted on mesoscale
devices~\cite{Rice1994,Wang2018} have shown the existence of a self-pulsing
regime which precedes the onset of coherent emission, where photon bursts are
recorded over time scales of the order of a nanosecond.  The self-pulsing
mechanism highlights the interaction between the slow energy reservoir (lasing
medium) and the fast-reacting photons in the pump range where fluctuations
play a crucial role.  However, fluctuations are not the sole cause of the
observations, since the nature of the interaction and the memory introduced by
the {\it slow} carriers~\cite{Dachner2010} are an essential ingredients which differentiate class
B (in this case semiconductor-based devices with low to moderate Q-factor) from the simpler class A
lasers~\cite{Tredicce1985} which are also expected to emit bursts of light, cf. Section~\ref{Aspikes}, with somewhat different features due to their Markovian nature. 

The purpose of this paper is to understand the generation of photon bursts observed in the emission of mesoscale lasers (where the fraction of spontaneous emission coupled into the lasing mode is $\beta \approx 10^{-4}$) below the
coherent emission threshold.  The present contribution focusses on the physical origin of pulses, which had been only observed in the experiment and in numerical simulations~\cite{Wang2015}, and on their description in terms of models.  As discussed throughout the paper, while correct temporal predictions require accounting for the intrinsic stochasticity of the interaction between inverted emitters and electromagnetic field, precious information arises from the topological structure of the radiation-matter problem which is well captured also by the Laser Rate Equations.  An additional point which is made here is that the non-instantaneous response of stimulated emission is an essential, intrinsic ingredient -- usually neglected in models due to its short characteristic time (by several orders of magnitude) compared to spontaneous processes -- indispensable in stochastic simulations.

\section{Background}

The central point of this paper is based on the experimental observation of bursts and on the interpretation of their origin.  In this section we introduce the necessary elements which lead to the observations, by giving a basic overview of the experiment (Section~\ref{expt}), and of the modeling tools which we use for the interpretation (Section~\ref{modelling}).

\subsection{Experiment and relevant observations}\label{expt}

\begin{figure}[htbp]
\centering
\includegraphics[width=\linewidth,clip=true]{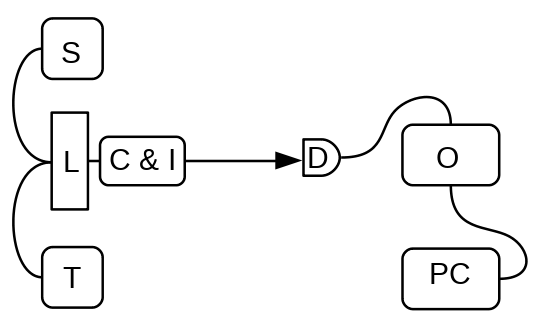}
\caption{
Experimental set-up.  S and T denote the current-stabilized power supply and the temperature controller, respectively.  L stands for the laser in its Thorlabs mount.  C \& I represent the collimator and isolation elements; the latter prevents re-injection of the emitted photons into the laser.  D stands for the detector, while O and PC denote the fast oscilloscope and the computer in which the data are stored.
}
\label{setup}
\end{figure}

The experimental set-up is shown in Fig.~\ref{setup} and consists of a commercial Vertical Cavity Surface Emitting Laser (VCSEL -- Thorlabs VCSEL-980) with nominal threshold current $i_{th,n} = 2.2 mA$ (maximum possible value $3 mA$~\cite{ThorlabsVCSEL980}), mounted on a TEC module (Thorlabs TCLDM9) which allows for temperature stabilization from an external source.  The estimated fraction of spontaneous emission coupled into the lasing mode is $\beta \approx 10^{-4}$ for this device (cf. Supplementary Information section in~\cite{Wang2015}). The nominal threshold is the one specified by the manufacturer as the minimum supplied current for correct, coherent operation of the device as a $2.5 \rm GHz$ telecommunication emitter.  The pump range we investigate is well below this nominal threshold.

The pump current is provided by a commercial DC power supply (Thorlabs LDC200VCSEL) with resolution 1 $\mu$A and accuracy $\pm$ 20 $\mu$A. The current drift at constant ambient temperature is less than $1 \mu$A over 30 minutes with low noise and ripple (both are $< 1.5 \mu A$ -- manufacturer's specifications~\cite{Thorlabs-ps}).  The active temperature stabilization comes from a home-built apparatus, capable of stabilizing to better than $0.1^{\circ}C$.  In the current range investigated, the laser emits on a single linear polarization, devoid of switching.

The signal output is passed through an isolation stage  and impinges onto a fast amplified photo-detector (Thorlabs PDA8GS, electrical bandwidth $9.5 \rm GHz$) through a multimode optical fibre (diameter $d_f = 62.5 \mu m$).  The electrical signal is digitized by a LeCroy wave master 8600A oscilloscope ($6 GHz$ analogue bandwidth, typically $1 \times 10^6$ sampled points).  All measurements are taken at a sampling rate $\tau_s^{-1} = 10 GHz$.  

The detector is supplied by a DC 12V battery, rather than by the power supply provided by the Manufacturer, to reduce the influence of electrical noise.  Special care is taken to minimize all external electrical noise sources, thus the leads carrying the DC current from the battery to the detector are shielded and grounded.  Ground loops are avoided in the detection section and a single high quality ground is used for the whole electrical acquisition chain.  Nonetheless, some residual electrical noise is present particularly with spectral components around 3\,GHz.

\begin{figure}[htbp]
  \centering
  \includegraphics[width=\linewidth,clip=true]{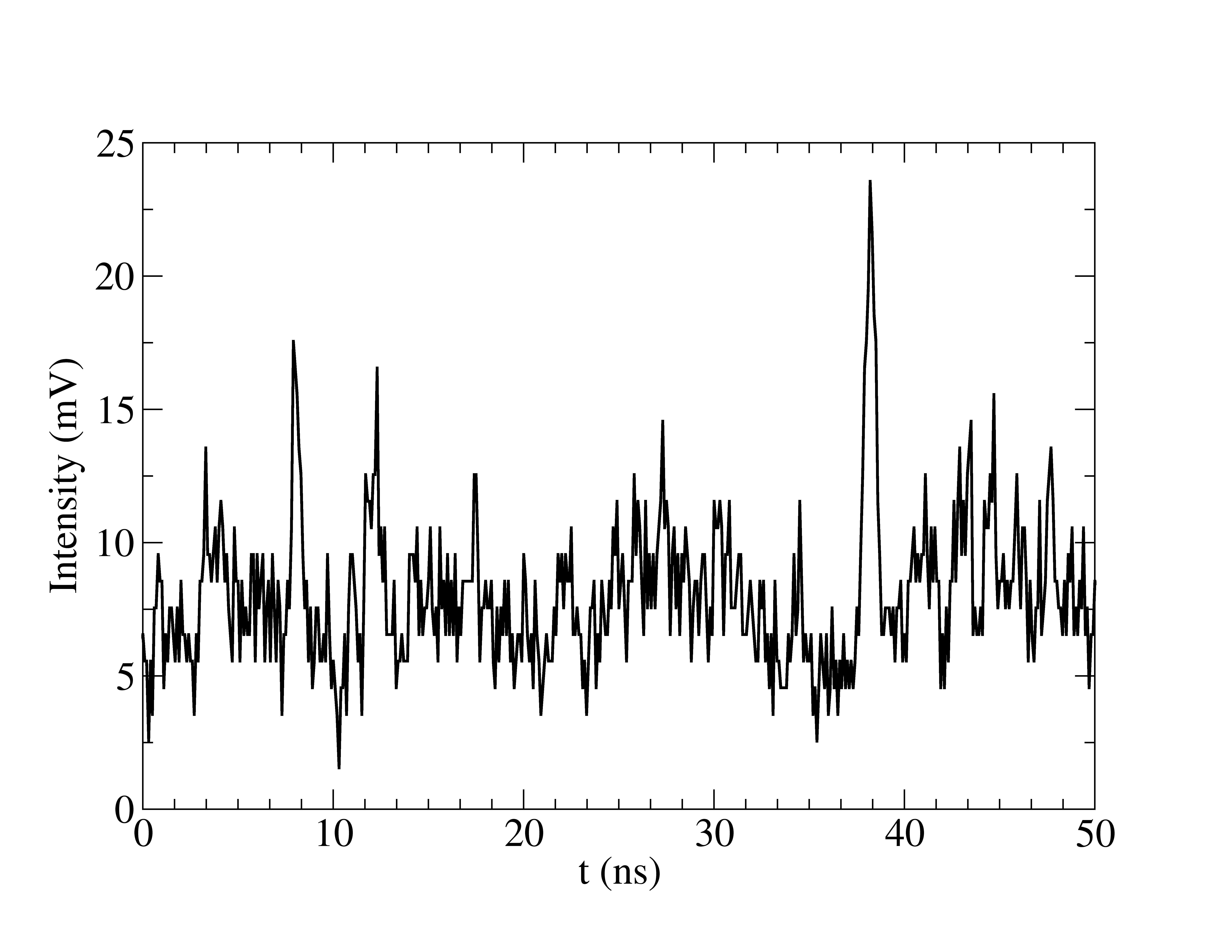}
\caption{
Temporal micro-VCSEL emission measured in the transition region between incoherent and coherent emission, at injection current $i = 1.26$ mA~\cite{Wang2015}.  A positive electrical offset, due to the detection apparatus, is present.
}
\label{spikes} 
\end{figure}

Measurements taken at the onset of emission show irregular photon bursts, with variable amplitude and temporal spacing (Fig.~\ref{spikes}); additional experimental details are available in~\cite{Wang2015}.  Similar phenomena have been observed in somewhat larger VCSELs with two polarized emission modes and have been attributed to competition between the two polarizations~\cite{Sondermann2003}, a phenomenon more recently observed in micro-pillars~\cite{Schlottman2018} and also in coupled photonic crystal cavities~\cite{Marconi2018}.  However, in the experiment from which the figure is taken, the second polarization mode is suppressed by at least 20dB, thus rendering the dynamics essentially single mode.

\subsection{Modelling}\label{modelling}

In the following we are going to use two different modelling approaches, one based on the standard Rate Equations (REs) which use a differential representation of the matter-field interaction~\cite{Coldren1995}, the other one fully stochastic, based on a recently introduced Stochastic Laser Simulator (SLS)~\cite{Puccioni2015}.  The latter is based on the semi-classical description of {\it emitter}-photon exchanges (where {\it emitter} stands for any two-level quantum source) described in Einstein's seminal paper on radiation theory~\cite{Einstein1917}.  In the SLS all processes are considered to be fully stochastic and the exchanges (photon creation, annihilation or transmission through the cavity mirror, as well as emitter dis/excitations) are counted by discrete integer numbers~\cite{Lebreton2013}.  The main features of this description, necessary for our discussion, are later highlighted in this subsection.

The REs, which have been successfully describing semiconductor laser dynamics for decades, describe the interaction between the real variables:  photon and carrier numbers.  The real-number representation is justified by the small variations which arise by the addition or subtraction, from the photon and carrier variables, of one (or a few) units compared to their respective large average values~\cite{Coldren1995}:
\begin{eqnarray}
\label{fre1}
\dot{n} & =  & -\Gamma_c n + \beta \gamma N (n+1) + F_{ph}(t)\, , \\
\label{fre2}
\dot{N} & = & R - \beta \gamma N n - \gamma N + F_{c}(t)\, .
\end{eqnarray}
Here $n$ and $N$ represent the photon and carrier number (or population inversion), respectively, $\Gamma_c$ and $\gamma$ are the relaxation rates for the intra-cavity
photons and for the population inversion, respectively, $R$ is the pump
rate and $\beta$ has been previously defined.  This set of REs includes the average contribution of the spontaneous emission to the number of coherent photons in the cavity mode through the term $\beta \gamma N$ (eq.~(\ref{fre1})).  The Langevin noise terms, reworked from~\cite{Coldren1995}, read:
\begin{eqnarray}
\label{photnoise}
F_{ph} & = &  \sqrt{2} \left( \sqrt{F_{nn}} G_a(0,1) + \sqrt{F_{nN}} G_b(0,1) \right) \, , \\
\label{popnoise}
F_c & = &  \sqrt{2} \left( \sqrt{F_{Nn}} G_c(0,1) + \sqrt{F_{NN}} G_d(0,1) \right) \, .
\end{eqnarray}
$G_j(0,1)$'s are independent Gaussian processes ($G_j \ne G_k$, $j \ne k$) with zero average and unity variance and
\begin{eqnarray}
F_{nn} & = & 2 \beta \gamma \overline{N} ( \overline{n} + 1 )\\
F_{nN} = F_{Nn} & = & - \beta \gamma \overline{N} \left( \overline{n} + 1 \right)\\
F_{NN} & = & R + \gamma \overline{N} + \beta \gamma \overline{N} \overline{n} 
\end{eqnarray}
are the $\delta$-correlated noise contributions~\cite{Coldren1995} computed following the McCumber approach~\cite{McCumber1966}.

The equilibrium values of the two variables, as directly obtained from eqs.~(\ref{fre1},\ref{fre2}), read:
\begin{eqnarray}
\label{nss}
\overline{n} & = &  \left\{ \left(\frac{C-1}{2}\right) + \sqrt{ \left(\frac{C-1}{2}\right)^2 + \beta C} \right\} \beta^{-1} \, , \\
\label{Nss}
\overline{N} & = & \frac{\Gamma_c}{\beta \gamma} \frac{C}{1 + \beta \overline{n}}\, , \\
\label{normP}
C & = & \frac{R}{R_{th}}\, , \\ 
R_{th} & = & \frac{\Gamma_c}{\beta} \, ,
\end{eqnarray}
where the over-strike represents equilibrium. 

The fully stochastic approach rests on the definition of the interaction processes as discrete recurrence relations where a probability of occurrence is assigned to each event with Poisson statistics.  Calling $M$ the number of excited emitters, $s$ the number of photons (present in the lasing mode) which stem from a stimulated emission process,  $r$ the photons on-axis (lasing mode) whose origin is a spontaneous relaxation and $r_o$ all other photons spontaneously emitted into all other electromagnetic cavity modes, we end up with the following recurrence relations~\cite{Puccioni2015}:
\begin{eqnarray}
\label{defN}
\label{MSS}
M_{q+1} & = & M_q + M_P - M_d - E_s \, , \\
\label{sSS}
s_{q+1} & = & s_q + E_s - L_s + s_{sp} \, , \\
\label{rSS}
r_{q+1} & = & r_{q} + D_L - L_r - s_{sp} \, , \\
\label{defRo}
r_{o,q+1} & = & r_{o,q} +(M_d - D_L) - L_o \, ,
\end{eqnarray}
where $M_P$ represents the pumping process (proportional to the pump rate),
$M_d$ the spontaneous relaxation processes, which reduce the population
inversion $M$, with branching into the lasing mode (represented by $D_L$) or
into all other electromagnetic cavity modes (collectively represented by
$M_d-D_L$), $E_s$ stands for the stimulated emission process which,
proportional to the product of $M_q$ and $s_q$, depletes $M$, $L_s$ and $L_r$
account for the leakage through the output coupler of stimulated and on-axis
spontaneous photons -- respectively -- probabilistically described through the
photon lifetime $\Gamma_c^{-1}$, and $s_{sp}$ describes the seeding process
which starts the first stimulated emission process, as explained in detail in
section~\ref{ignition}.  $L_o$ represents the losses for the off-axis fraction
of the spontaneous photons ({\it laterally}) exiting the cavity volume with
lifetime $\Gamma_o^{-1}$, similarly to $L_{s,r}$.  All processes are
probabilistic and follow Poisson statistics. 

Technical details on the numerical implementation of the scheme can be found
in~\cite{Puccioni2015}, where it is noted that Poisson processes can be
approximated as Binomial Distributions in the limit of small arguments
(necessary also for the additivity of the distributions), often with some
advantages in computing time.  The time step over which each process is
estimated needs to be adjusted to satisfy additivity, i.e., the argument must
remain sufficiently small (in practice of the order of $0.1$), thus the
numerical scheme must be accordingly organized.  A time $\tau = 2 \times
10^{-15} s$ is a good starting choice.

The advantage of the fully stochastic scheme is that it naturally and properly accounts for the discreteness of each interaction and for its intrinsic randomness, unlike the RE approach, which is typically based on a Langevin noise description, with the accompanying hypotheses and shortfalls~\cite{Lippi2018,Lippi2019}.  It is, however, important to stress the fact that both descriptions are based on the same fundamental physical processes.  While the REs can be derived from Maxwell-Bloch equations when the polarization variable can be adiabatically eliminated~\cite{Narducci1988}, they can also be phenomenologically written~\cite{Coldren1995} starting from Einstein's semi-classical foundations~\cite{Einstein1917}.  As a result, the two approaches share the same basic description, differing only in the way of describing some details of the interaction.  In other words, the phase space topology~\cite{Lippi2000a,Lippi2000b} is the same for the two models and the information that it provides holds for the interpretation of both (cf. section~\ref{topology}).  The same understanding about the common origin of the two modeling approaches explains the validity of analytical predictions obtained from the REs (e.g., in the statistical indicators) in agreement with the numerical results obtained from the SLS~\cite{Lippi2019}, while the numerics obtained directly from the REs fail.

\section{Experimental observations and model predictions}

\begin{figure*}[htbp]
  \centering
  \includegraphics[width=0.45\linewidth,clip=true]{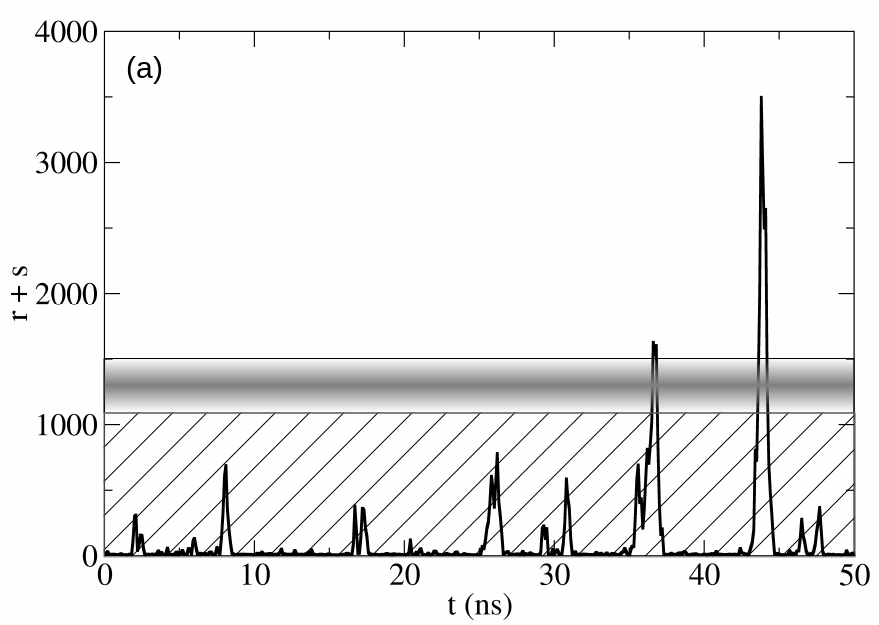}
    \includegraphics[width=0.45\linewidth,clip=true]{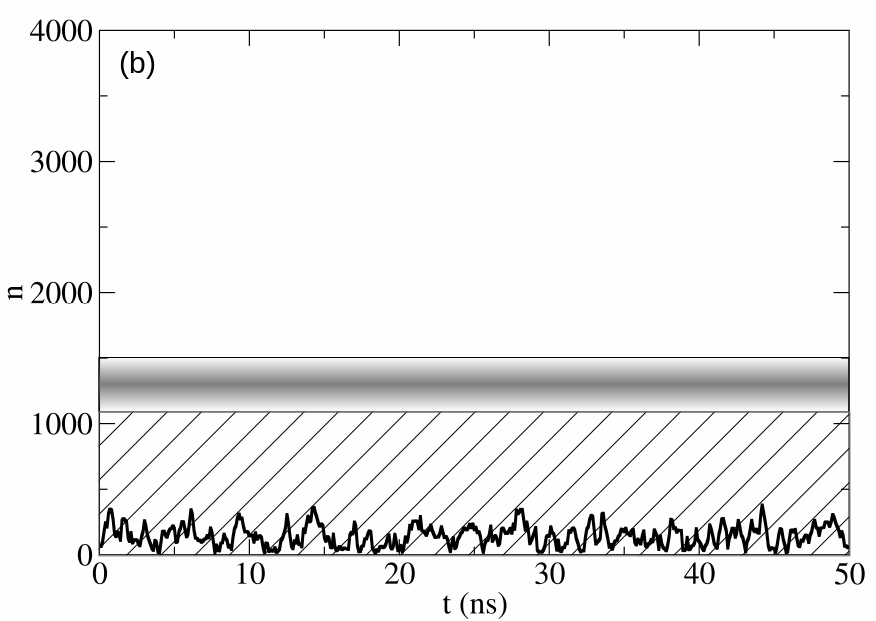}
\caption{
(a):  total photon output (spontaneous, $r$, plus stimulated, $s$) predicted by the SLS for pump equal to its threshold value, $C = 1$ (eq.~(\ref{normP})).  (b):  total photon output ($n$) predicted by the REs for $C = 1$.  The shaded area corresponds to the detection limit (darker central line) and covers the electronic noise interval which corresponds to the slow oscillations of Fig.~\ref{spikes}:  the area where background noise oscillates and reveals possible small peaks.  The dashed area below the band represents the photon range which is entirely masked by noise.  The entirely different temporal trajectories imply that while the photon bursts predicted by the SLS are compatible with experimental observations (cf. Fig.~\ref{spikes}), the photon flux computed by the REs is never observable.  Here and in the following figures we report intracavity photon numbers.
}
\label{numspikes} 
\end{figure*}

Fig.~\ref{numspikes} compares the expected laser emission computed with two different schemes.  Panel (a) shows the photon flux predicted by the SLS using a detector with bandpass equivalent to that of the experiment, for a pump value equal to the nominal threshold, which satisfactorily matches the experimental result of Fig.~\ref{spikes}.  The emission takes the form of large pulses over an extremely low background mostly constituted of spontaneous photons.  The horizontal darker line corresponds to the minimal photon flux which can be detected by the experimental apparatus:  the detector is not sufficiently sensitive to give a visible signal for photon number values below this line.  The shaded area corresponds to the electrical noise band of the detection chain, including noise and long-terms fluctuations:  signals below this area are entirely masked, while those whose peak falls into the band can be at times recognized, depending on the details of the instantaneous detection noise.

Fig.~\ref{numspikes}b shows the photon flux predicted by the REs with Langevin noise (eqs.~(\ref{fre1},\ref{fre2})), for the same pump value used for the SLS.  The scale is chosen as in Fig.~\ref{numspikes}a to allow for a direct comparison of the predictions:  the signal is well below the detection limits and does not show the features (peaks) observed in Fig.~\ref{spikes}.  It is important to notice that the averages for the two numerical predictions are compatible with each other, since $\langle r+s \rangle = (0.1 \pm 0.3) \times 10^3$ photons from the SLS, while $\langle n \rangle = (0.14 \pm 0.08) \times 10^3$ photons computed from the REs.  On the other hand, the relative standard deviations are quite different: $\frac{\sigma_{SS}}{\langle r+s \rangle} = 3$, while $\frac{\sigma_{RE}}{\langle n_{RE} \rangle} \approx 0.6$, in agreement with the visual information.  The prediction from the SLS gives extremely large spikes over a negligible background, while the one from the REs correspond to a standard noisy signal.  It is also interesting to notice that the average power corresponding to the average photon flux (here computed on $\Delta t = 0.1 ns$, in agreement with the experimental detection bandwidth) is $P \approx 0.1 \mu W$, one order of magnitude below the experimental detection capabilities, while the peaks observed in Fig.~\ref{numspikes}a, up to 30 times larger, are perfectly compatible with the observations (cf. Fig.~\ref{spikes}).
While the previous discussion analyses the results obtained from devices with a specific cavity volume (corresponding to $\beta \approx 10^{-4}$), similar observations have been obtained with larger $\beta$ values~\cite{Wang2017a}, thus proving the generic nature of the behaviour.

\vspace{1cm}

\subsection{Contradictions}\label{contradictions}

The reported behaviour jars with several aspects of the commonly accepted
laser features, partly due to modeling issues, which we later examine, partly
because several fundamental laser characteristics have been studied for
devices which belong to the dynamical Class A~\cite{Arecchi1984,Tredicce1985}.
The transition through threshold has been thoroughly investigated
experimentally in He-Ne and dye
lasers~\cite{Arecchi1966,Arecchi1967,Arecchi1971,Roy1980,Kaminishi1981}, both
macroscopic Class A devices.  The main finding is that the transition takes
place through steps which lead from a predominantly incoherent to an,
eventually, fully coherent emission, representable as a {\it statistical
mixture} of spontaneous and stimulated photons whose average number progresses
towards a predominance of the latter species as the device approaches
lasing~\cite{Arecchi1967V}.  

The laser transition in class A macroscopic devices can be assimilated to a
thermodynamic phase transition~\cite{Degiorgio1970,Dohm1972}, but as the
cavity volume decreases the nature of the transition changes even in the
simplest (class A) devices~\cite{Takemura2019}, with the $\beta = 1$ limit
presenting perfect coherent emission for all pump values:  the threshold has
entirely disappeared.  This feature corresponds to the {\it threshold-less
laser} whose existence had been postulated long
ago~\cite{Yokoyama1989,Bjoerk1991} and which has opened many questions on the
transition to coherence in Class B
devices~\cite{Bjoerk1994,Rice1994,Ning2013}, dominated by semiconductor-based
systems where the slower carriers maintain memory of the temporal evolution
over the faster-relaxing photons.  For this class of systems, to which the
devices used in our current investigation belong, the statistical properties
of the radiation had been originally studied at the macroscopic
scale~\cite{Paoli1988,Arecchi1989,Ogawa1990}, later leading to investigations
in solid-state micro-devices~\cite{vanDruten2000,Lien2001,Woerdman2001} whose
behaviour was found in excellent agreement with the macroscopic theoretical
predictions.

Aside from later statistical measurements performed in semiconductor-based devices~\cite{Strauf2006,Ulrich2007} which could only provide statistical information and where hints of a more complex scenario may be detected, the detailed investigations~\cite{vanDruten2000,Lien2001,Woerdman2001} seemed to conclusively prove that the transition from incoherent to coherent
emission appears to occur in agreement with known models, which exclude the
possibility for the dynamical photon bursts on which we report here.  These
bursts are responsible, among others, for the failure of the macroscopic class
B photon statistics in describing the threshold transition~\cite{Wang2017a}
and are incompatible with the RE description, as shown in
Fig.~\ref{numspikes}.  What is therefore the origin of these spontaneous
photon spikes?

In the double-logarithmic representation of the average photon number emitted
by the laser as a function of pump rate (see, as an example,
Fig.~\ref{scurve}, discussed in section~\ref{ignition}), the region connecting
the below-threshold straight-line response to the corresponding one above
threshold is commonly called the Amplified Spontaneous Emission (ASE) region
thanks to a somewhat loose visual analogy of the observations
(Fig.~\ref{spikes}) with those which stem from ASE~\cite{Hayenga2016,Pan2016}.
In spite of these similarities, the analogy is however misleading.  ASE
requires very large gain factors ($O(10^3) cm^{-1}$) which are matched in
Quantum Well-based VCSELs (gain is estimated at $\approx 10^3
cm^{-1}$~\cite{Coldren1995}), but it also takes place over nonnegligible
propagation distances.  It is not uncommon for the latter to be of the order
of a few centimetres (in solids) or meters (in gases).  The gain region in a
VCSEL is $\lesssim 10 nm$, thereby leading to negligible single-pass
amplification.  The very low single-pass amplification is indeed the reason
for the need for high reflectivity Bragg mirrors, which recycle intra-cavity
photons up to $O(10^3)$ times before the photon's escape probability from the
cavity becomes sizeable.  Since the same medium is repeatedly visited by the
circulating photons during their multiple cavity round-trips, it is not
possible to assimilate the pulse formation to the one observed in ASE, where
the pulse amplification stems from the full and instantaneous depletion of the
local inversion by the advancing optical field. 

A more suitable analogy to describe this regime is laser gain-switching (or, more properly, pump-switching).  However, at this stage this remains a simple pictorial analogy since the observation of photon bursts cannot be due to changes in the pump, since the short-term pump stability is $\Delta i \ll 10^{-2} i$ in our experiment (\cite{Wang2015} and in its accompanying Supplementary Information section).   Nonetheless, a simplified analysis based on REs can show that {\bf \em if} the carrier number is temporarily driven above the threshold value {\bf \em then} a pulse will ensue, and that it is possible to predict both the typical duration of a pulse and a maximum repetition rate (devoid of regularity, in agreement with the experiment) with quantitative values in good agreement with the observations~\cite{Wang2017}.  The analysis of section~\ref{topology} gives a sounder foundation to these considerations and supports the more detailed stochastic predictions.

\section{Topological information}\label{topology}

Even though the REs cannot predict the dynamics of small lasers, the information they can provide about the topology of the phase space holds for devices at all scales (as is the case for the analytically derived noise properties~\cite{Lippi2019}).  We can therefore resort to the stability analysis in the region of interest, i.e., $C < 1$, setting

\begin{eqnarray}
\frac{d}{d t} 
\left(
\begin{array}{c}
\nu \\
\mu \\
\end{array} \right) & = & \mathcal{\overline{\overline{S}}} \quad \left(
\begin{array}{c}
\nu\\
\mu\\
\end{array} \right) \, \\
\mathcal{\overline{\overline{S}}} & = & 
\left( 
\begin{array}{ c c }
\beta \gamma \overline{N} - \Gamma_c - \lambda & \beta \gamma (\overline{n} + 1) \\
-\beta \gamma \overline{N} & -\gamma - \beta \gamma \overline{n} - \lambda \\
\end{array} \right)
\end{eqnarray}
where we have defined the perturbations
\begin{eqnarray}
n(t) & = & \overline{n} + \nu e^{\lambda t} \, , \\
N(t) & = & \overline{N} + \mu e^{\lambda t} \, ,
\end{eqnarray}
with the steady states for the photon and carrier number denoted by the overlines.

\begin{figure}[htbp]
  \centering
  \includegraphics[width=\linewidth,clip=true]{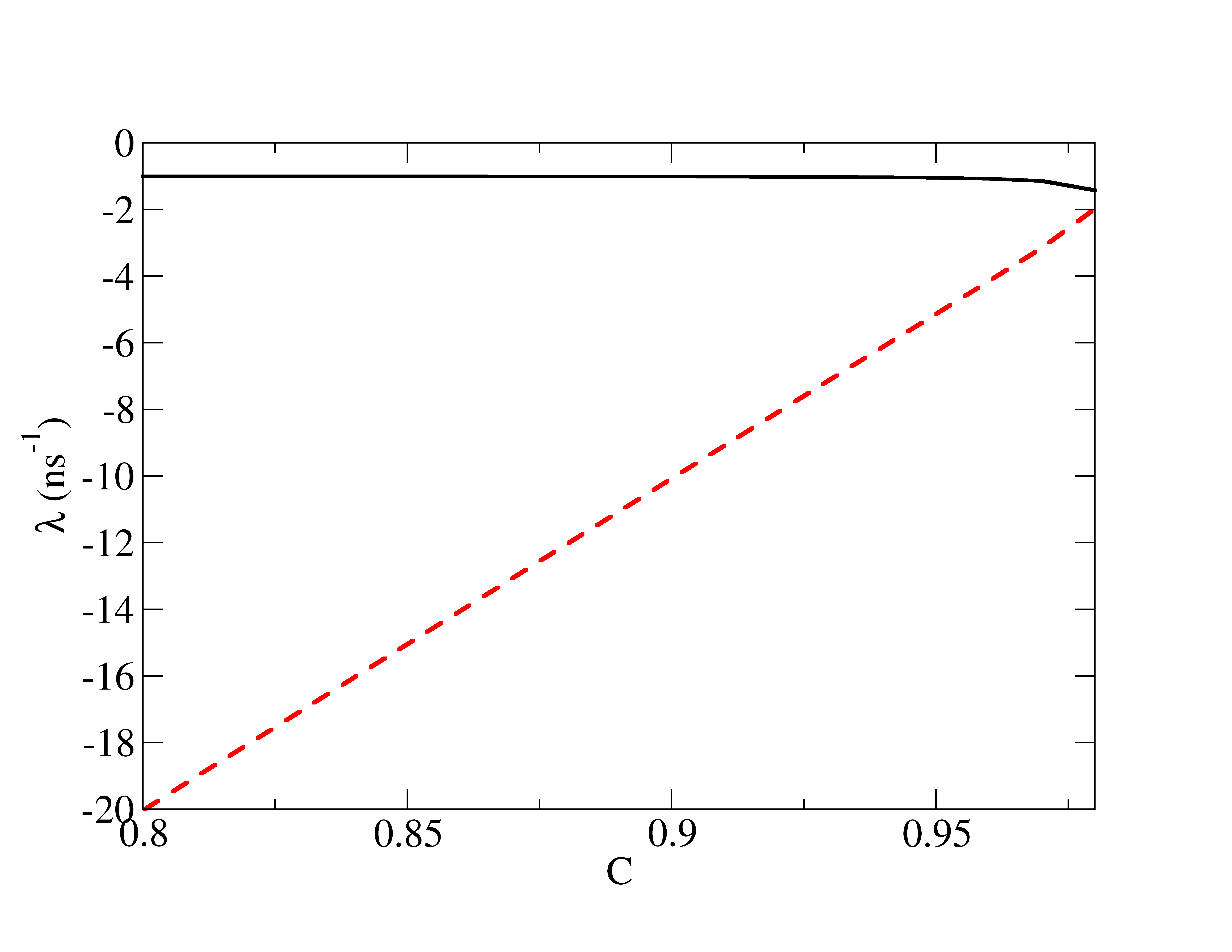}
\caption{
Eigenvalues of the REs for $\gamma = 10^9 s^{-1}$, $\Gamma_c = 10^{11} s^{-1}$ and $\beta = 10^{-4}$ in the below-threshold region as a function of normalized pump $C$:  the solid line (black online) corresponds to $\lambda_1$, the dashed one (red online) to $\lambda_2$.
}
\label{eigenval} 
\end{figure}

The eigenvalues in the region $0.8 \le C \le 0.98$ are shown in Fig.~\ref{eigenval}.  Both are negative, indicating that the below-threshold solution is stable, but the contraction is much stronger (rate $\lambda_2 \ll 0$) along one directly than in the other, since $\lambda_1$ remains all the time much closer to $0$.  In addition, the eigenvalues are real, thus showing that relaxation along the two eigendirections occurs monotonically.

\begin{figure}[htbp]
  \centering
  \includegraphics[width=\linewidth,clip=true]{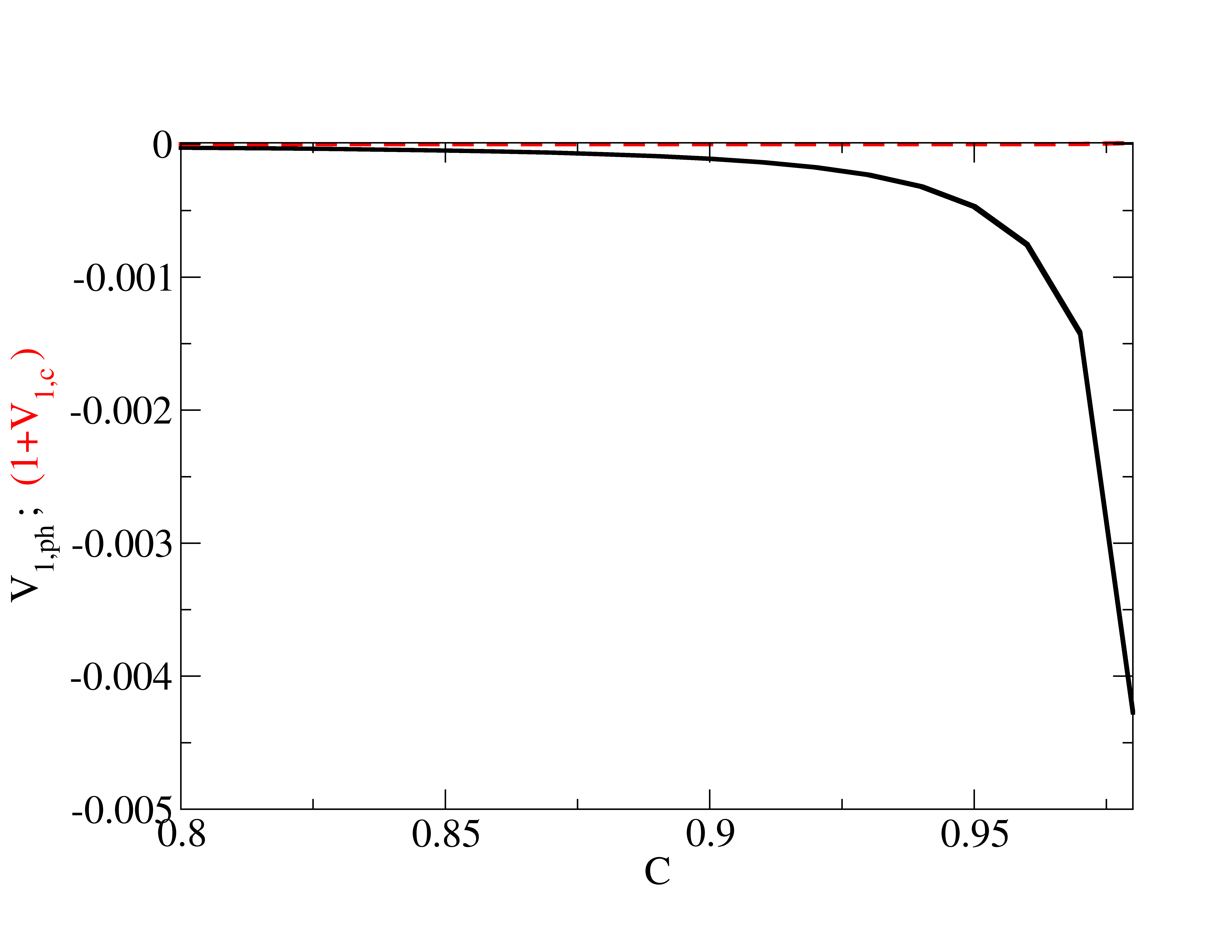}
\caption{
Eigenvector components of $V_1$ (corresponding to $\lambda_1$):  solid line (black online) represents the amplitude of the {\it photon} component, while the dashed line (red online) that of the {\it carrier} one, to which $1$ has been added:  this component is very close to $-1$ over the whole displayed pump interval.   
}
\label{evecl} 
\end{figure}

The eigenvectors identify the directions along which the contracting dynamics, identified by the two eigenvalues, develops.  Fig.~\ref{evecl} shows the evolution of the two normalized components of the eigenvector $V_1$ corresponding to $\lambda_1$:  the solid line (black online) shows the {\it photon} component of the eigenvector, while the dashed one (red online) gives the {\it carrier} component to which  $1$ has been added (it would be too close to $-1$ to display both components on the same graph with enough resolution).  This eigenvector is practically entirely aligned along the {\it carrier} variable $N$ in the negative direction, while the {\it photon} component remains negligible even at the largest pump value $V_{1,ph}(C = 0.98) \approx -0.0043$.  Thus, we conclude that the slow contraction (rate $\lambda_1$) occurs (almost) exclusively in the direction of the carriers variable in the plane.  In other words, a perturbation of the carriers will slowly relax towards its equilibrium value without affecting the photon number.

\begin{figure}[htbp]
  \centering
  \includegraphics[width=\linewidth,clip=true]{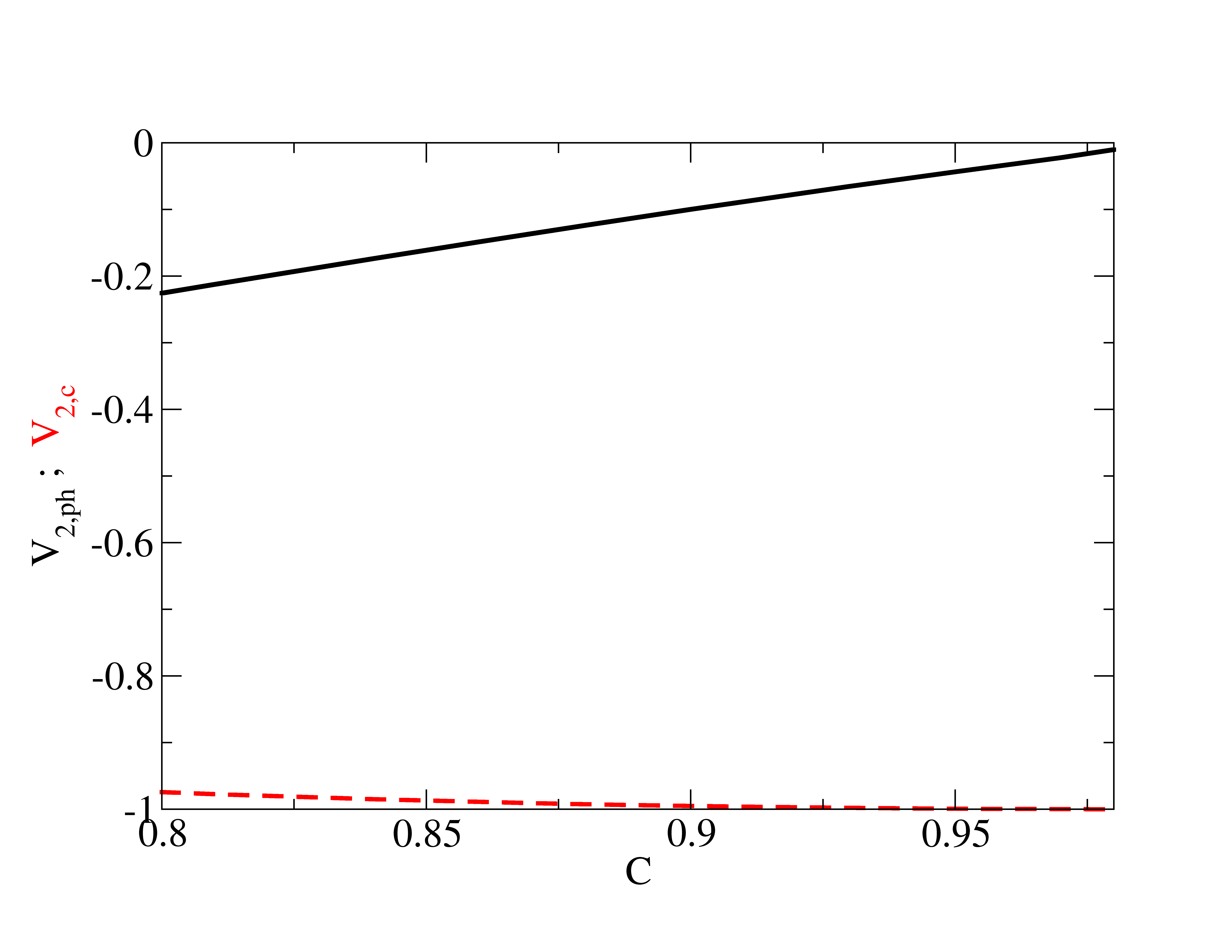}
\caption{
Eigenvector components of $V_2$ (corresponding to $\lambda_2$):  the solid line (black online) represents the amplitude of the {\it photon} component, while the dashed line (red online) that of the {\it carrier} one.   The amplitude of the {\it photon} component at $C = 0.98$ is approximately a factor $2$ larger (in modulus) than the one of $V_1$; thus -- even though larger than for $V_1$ --, it remains negligible compared to the {\it carrier} component close to threshold.
}
\label{evecm} 
\end{figure}

Fig.~\ref{evecm} shows the equivalent components for the more strongly contracting eigendirection, with the same conventions for the lines.  Sufficiently far from threshold ($C = 0.8$) the {\it photon} component of $V_2$ amounts to approximately 20\% of the amplitude, but the ratio becomes only slightly larger than the one of $V_1$  (by only a factor 2) at $C = 0.98$.  

Eigenvalues and eigenvectors give precious indications for the description of threshold dynamics.  {\it Far below threshold} (i.e., here $C = 0.8$) one eigenvalue, $\lambda_2$, relaxes very quickly to its equilibrium solution (twenty times faster than $\lambda_1$ and only five times slower than the cavity decay rate $\Gamma_c$).  At this pump value the contracting direction is oriented for 20\% in the {\it photon} component, in other words, $V_2$ forms an angle $\approx 0.2 rad$ with the $N$ variable.  This angle becomes almost entirely negligible ($O(10^{-2}) rad$) close to threshold.  For $V_1$ the alignment with the {\it carrier} direction is nearly stable throughout the whole pump interval, the largest deviation occurring at $C = 0.98$ when the angle with $N$ is $O(10^{-3}) rad$.  

The consequence of this analysis is that the eigendirections are mostly aligned with the {\it carrier} direction over the whole pump interval, below threshold, while the photon direction plays only a minor role.  This is physically understandable, since below threshold the only photon contribution comes from the spontaneous emission, while the energy is always stored in the carrier variable $N$.  In particular, the least stable direction $V_1$, corresponding to the rate $\lambda_1$, is ``strictly'' aligned with the carriers.  Thus, if we picture a fluctuation in the carrier number away from threshold, it will converge back relatively rapidly along $V_2$ (where a {\it photon} component exists) and then will continue evolving slowly along $V_1$ without any consequence on the photon number $n$.  This picture translates the fact that the carriers are the slow variable which carry the memory throughout the time evolution, while photons reacts very quickly.  Above threshold, i.e., when the two variables are coupled, there is an interchange of energy which manifests itself in the form of the so-called {\it relaxation oscillations}~\cite{Coldren1995}, but below threshold this mechanism is missing due to the lack of stimulated emission:  the photon number hardly feels the consequence of a fluctuation.

Below threshold the carrier number in a Quantum Well laser is still sizeable; for our device with $\beta \approx 10^{-4}$ and normalized pump $0.8 \lesssim C \lesssim 1$ it can be estimated from the REs (eq.~(\ref{Nss})) to be of the order of $10^6$.  Assuming Poisson statistics, fluctuations $\Delta N \lesssim 10^3$ are common.  Naively, one would expect to find an equivalent amount of photon fluctuations, which, instead, have a much smaller amplitude, even though their relative contribution is much larger than the average photon number $\left( \frac{\langle \Delta n \rangle}{\langle n \rangle} > 1 \right.$, contrary to what happens for the carrier fluctuations where $\left. \frac{\langle \Delta N \rangle}{\langle N \rangle} \ll 1 \right)$.  The preceding topological considerations help understand this fact:  the nearly perfect alignment of the most unstable eigendirection with its {\it carrier} component implies that there is only an extremely small projection ($\lesssim 10^{-3}$) -- which can be translated into a corresponding probability -- onto the {\it photon} component.  In other words, fluctuations in the carrier number can occur without almost any influence onto the photon number.

If we now consider the slow dynamics of the carrier number, we find that it is possible to accumulate a large excess of carriers, due to the unlikely transfer of the fluctuation onto the photon number.  However, {\it when a large carrier excess takes place {\bf \em and} coupling to the radiation field occurs}, then the conditions match those normally encountered in the so-called gain-switching, where a large excess of population is suddenly discharged into the photon field~\cite{Lau1988}.  At variance with the driven switching process (e.g., through pump modulation), where the amplitude of the fluctuation can be directly related to the delay time for pulse buildup~\cite{Balle1994,Grassi1994}, the process in this case is spontaneous.  The common point, however, is the appearance of photon bursts emitted from the cavity, in place of a continuous, albeit noisy, flow predicted by the REs and expected from the traditional threshold considerations.

From an analysis based on REs it is, however, possible to obtain estimates of the buildup times to be expected.  By imposing a steady growth in the carrier number $N$ through an above-threshold pump fluctuation, it is possible to obtain, through suitably approximated analytical calculations, an estimate of the shortest possible rate at which spontaneous pulsing may occur~\cite{Wang2017}.  Comparison with experimentally measured radio-frequency (rf) power spectra in devices of the kind discussed here shows that the high-frequency spectral cutoff is quantitative agreement with the estimated rates~\cite{Wang2017}.  Thus, from the REs we can obtain several quantitative indicators which illustrate the basic physical phenomena responsible for the photon bursts, once the physical mechanisms are postulated.  This is the first set of missing physical elements explaining the failure to predict spikes which compound the improper description of noise (only {\it high-frequency}~\cite{Lippi2018} and {\it small-amplitude} Langevin components), instead of properly accounting for the discreteness of the emission and interaction processes~\cite{Lippi2019}.  An additional, crucial, element will be discussed in section~\ref{ignition}.

We can close this section by remarking that the {\it mixing} that the RE description operates on the {\it origin} of photons (thus on their degree of coherence) requires an effort of interpretation in handling the information that is derived from it.  For instance, the eigenvector analysis, which has just shown its usefulness in explaining the possible origin of photon bursts, mixes the spontaneous, on-axis photons which correspond to the eigenvalues (and eigenvectors) with the {\it coherent} photons which are emitted in a burst.  The weakness of the coupling of the electromagnetic field below threshold corresponds to the fact that spontaneous photons exert negligible influence onto the energy reservoir (carriers), since they simply play the role of a {\it rejection channel} for the energy stored in the population.  The coupling strength immediately changes as soon as stimulated emission takes over (in the bilinear term of the REs -- eqs.(\ref{fre1},\ref{fre2}) -- or in the $E_s$ term of the SLS -- eqs.~(\ref{MSS},\ref{sSS})).  Seen in this light, the abrupt change in behaviour is no longer surprising, since it corresponds to two different kinds of emissions.  However, only the SLS makes the distinction between the two {\it emission channels} while the RE analysis blurries the picture by treating the two kinds of photons as being of the same sort because they belong to the same mode (but have different coherence properties).

\section{``Ignition'' of stimulated emission}\label{ignition}

The central point in the explanation of the experimental observations, and the main difference between the numerical integration of the REs and of the SLS, is the start of stimulated emission and a correct, instantaneous description of its amplification.  REs provide useful information on the phase space structure and describe the average stimulated emission, but fail to produce a satisfactory temporal description of its evolution (cf. Fig.~\ref{numspikes} and~\cite{Lippi2019}). 

\begin{figure}[htbp]
  \centering
  \includegraphics[width=\linewidth,clip=true]{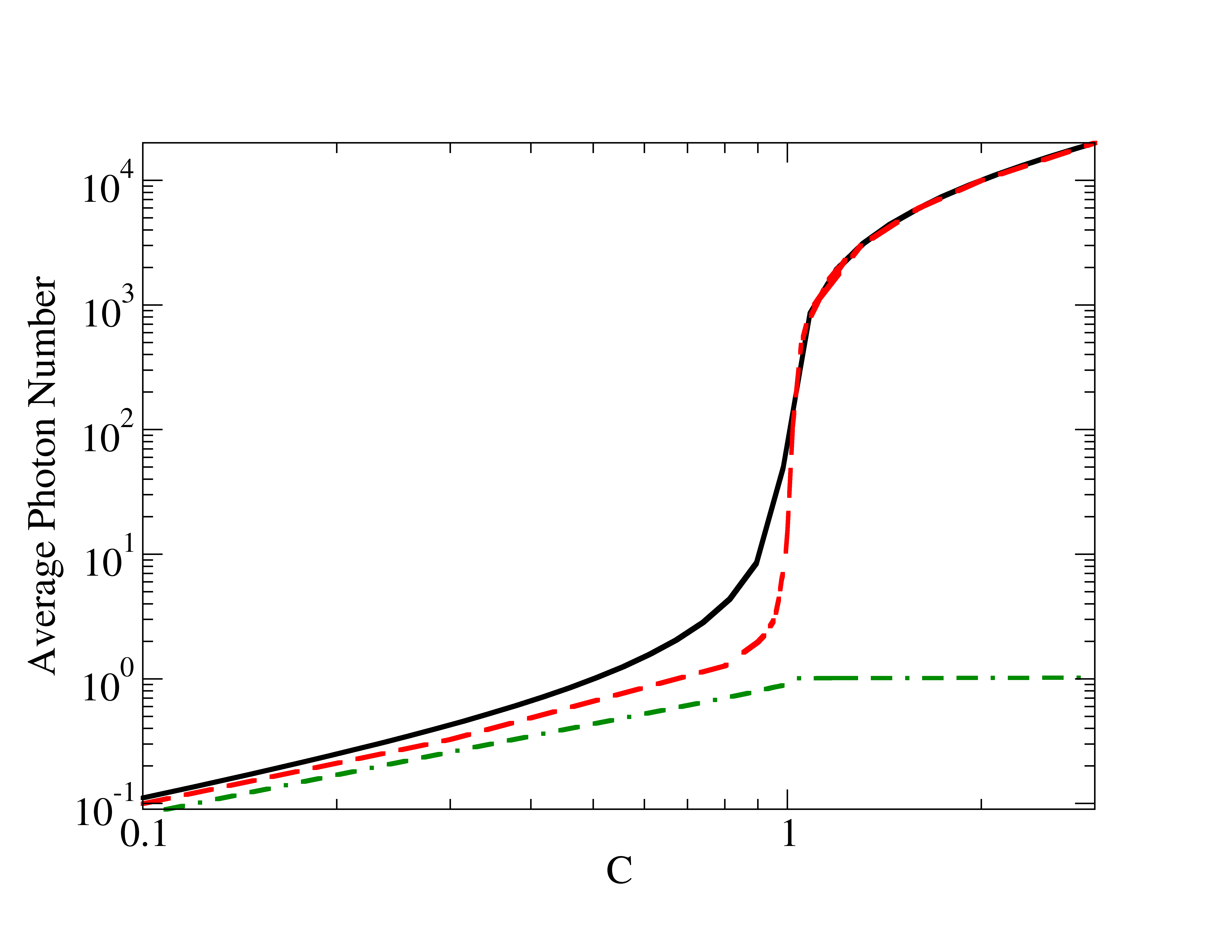}
\caption{
Average photon number emitted by a $\beta = 10^{-4}$ laser, as a function of pump rate $C$ (normalized to its threshold value, eq.~(\ref{normP})), as computed with REs (solid, black online) or through the SLS (dashed, red online -- this trace represents the total photon number emitted, i.e., $(r+s)$, cf. eqs.~(\ref{sSS},\ref{rSS})).  The double-dash--dotted line (green online) represents the number of spontaneous photons ($r$ from eq.~(\ref{rSS})).  Computation parameters:  $\gamma = 3 \times 10^9 \rm s^{-1}$, $\Gamma_c = 10^{11} \rm s^{-1}$.
}
\label{scurve} 
\end{figure}

A direct comparison between the average photon number predicted by the REs (solid line) and the total photon number (stimulated $s$ plus spontaneous $r$) predicted by the SLS (dashed line) is shown in Fig.~\ref{scurve} in double-logarithmic scale as a function of normalized pump rate $C$.  
The most immediate remark is that the agreement between the two methods is excellent above threshold -- already from the upper third of the steeper part of the curve --, while the two curves converge again only at very low pump rates (left part of the $C$-axis).  Instead, in the intermediate, below-threshold region the REs predict a larger average number of photons present in the laser.  The double-dashed--dotted line shows the sole contribution of the spontaneous emission ($r$ variable, eq.~(\ref{rSS})) computed by the SLS.  Comparison of the total photon number (dashed line) to the spontaneous one (double-dashed--dotted line) shows that the contribution of stimulated emission remains negligible until $C \approx 0.98$.

A crucial point in interpreting the laser response predicted by the REs rests in the correct interpretation of its average photon number.  While for the SLS two distinct variables account for the number of photons present in the lasing mode ($r$ for the spontaneously emitted ones, $s$ for the those resulting from a stimulated process), the REs make no such distinction:  $n$ represents the number of photons on-axis without any consideration of their nature (thus, indirectly, of their phase coherence, although neither approach considers phase).   As a consequence, the lower branch of the s-shaped curve of Fig.~\ref{scurve} -- sufficiently far from the steep growth -- is made up predominantly of spontaneous photons (solid line).  Similarly, the upper branch represents for the largest part the average number of stimulated photons.  However, approaching threshold from below and in the steeper parts of the curve connecting the two main branches, the average photon number $\langle n \rangle$ is a mixture of spontaneous and stimulated photons (solid line) with variable proportions depending on the pump rate.  In this sense, the RE picture matches the description of threshold as 
a superposition (in average) of spontaneous and stimulated photons.  This description, however, holds only for the average quantities and cannot account for the start of the amplification process.  Since the latter is central to the interpretation of the observations (Fig.~\ref{spikes}), we focus the rest of this section on the start of the stimulated process.

Stimulated emission typically occurs at a much faster rate than its spontaneous counterpart and therefore is often considered as being ``instantaneous''.  However, when considering the {\it ignition}  of stimulated emission, starting from an initial photon present in the lasing mode, a realistic picture requires an explicit account of the time needed for this process to occur.  In other words, we need to describe the mechanism -- and find a way of reproducing it numerically -- which starts populating the $s$ variable in the SLS (eq.~(\ref{sSS})), subsequently amplified by $E_s = \beta \gamma s M \tau$, where $\tau$ is the time interval over which the probabilities are computed in the SLS scheme. 

An estimate of the rate of stimulated emission requires the computation of the transition probability in terms of Fermi's golden rule~\cite{CohenTannoudji}, taking into account the semiconductor bands and the electromagnetic cavity's density of states as a function of frequency.  This computation, laborious and requiring suitable approximations, is explicitly done in~\cite{Coldren1995} (in particular in Chapter 4 and Appendix 6) delivering a simplified and very practical form for quantitative estimates:
\begin{eqnarray}
R_{st} & = & v_g g \mathcal{N} \, .
\end{eqnarray}
Here $v_g$ stands for the group velocity, $g$ for the gain and $\mathcal{N}$ for the number of photons present in the mode.  The specific case of VCSELs operating at $\lambda = 980 nm$ is considered in~\cite{Coldren1995}.

Quantitative estimates of $R_{st}$, necessary for the numerical computations, require detailed knowledge of the structure of the device -- an impossible task with a commercial device.  Proceeding with some judicious choices based on the detailed information offered in~\cite{Coldren1995}, we can arrive at $v_g \approx 3 \times 10^8 m s^{-1}$ and $g \approx 3 \times 10^5 m^{-1}$ for the injection current interval corresponding to the start of stimulated emission
\footnote{Assuming an active diameter for our micro-VCSEL $d \approx 6 \mu$m (cf. Supplementary Information in~\cite{Wang2015}), for an injection current $i \approx 1$ mA the corresponding current density is $J \approx 40 J_{tr}$ (taking the order of magnitude $J_{tr} \approx 10^2$ A/cm$^2$, second to last column, from Table 4.5 of~\cite{Coldren1995} for a GaAs device:  $J_{tr}$ transparency current density).  Given the lack of detailed construction information for our device, we also keep only the order of magnitude for $\frac{J}{J_{tr}} \approx 10$.  From the same table, the value for $g_0 \approx 10^3$ matching this class of devices (restricted again to the order of magnitude) provides the estimate for $g$ given in the text.  Since we are interested in highlighting the physical principles of the role played by the characteristic time of the stimulated emission, rather than in accurately computing its value -- cf. discussion later in this section --, the order of magnitude provides sufficient information.  In the same way, the value for the group velocity $v_g$ (Table 5.1 in~\cite{Coldren1995}) needs to be accurate only to the order of magnitude for our present purposes.}
, thus obtaining
\begin{eqnarray}
R_{st} \approx 9 \times 10^{13} s^{-1} \, ,
\end{eqnarray}
where we have used $\mathcal{N} = 1$, since it matches the {\it ignition} of stimulated emission starting from {\it one spontaneous photon} in the lasing mode.  
For a time increment $\tau$ the probability of obtaining a stimulated emission event, in the presence of one spontaneous photon in the lasing mode, is therefore $\mathcal{P}(\tau) = \tau R_{st}$ which for the value of $\tau$ typically used in the SLS simulations, in the threshold region, corresponds to $\mathcal{P} \approx 0.18$.  While the analytical procedure of~\cite{Coldren1995} is entirely valid, the numerical estimate of $\mathcal{P}$ has to be taken with some precaution, because of the above considerations.  We can nonetheless consider it as a good starting point for our conceptual analysis.

This reasoning explains the use of the $s_{sp}$ term in eqs.~(\ref{sSS},\ref{rSS}) which initiates the population of the $s$-class (stimulated photons):  when $s_q=0$ and $r_q>0$, then {\it one} photon is removed from the $r$-class (spontaneous photons) and is promoted to the $s$-class with (Poissonian) probability $\mathcal{P}(\tau) \times r$.  This scheme allows for an acceptable reproduction of the physical process which starts populating the $s$-class which would otherwise remain forever unpopulated (for $s=0$) since the stimulated process, $E_s \propto s$, requires $s > 0$ in order to take place.

This construct is at the heart of the difference between the RE and the SLS representation of laser operation -- in addition to the use of random processes and integer variables -- and plays a crucial role not only at the transition between incoherent and coherent emission, but also in the pump rate range where lasing is not sustained (even if $\langle s \rangle > 0$).  The average superposition of incoherent and coherent photons, which characterizes the transition between the {\it below threshold} and {\it above threshold} regimes of emission, is now consistently described with two variables whose averages evolve according to the {\it average} physical processes as described by the differential model (REs) but with temporal evolution dictated by the randomness of the probabilistic realization of each kind of event (pump, spontaneous or stimulated emission and transmission through the output mirror).  

As a last remark, we now discuss in more detail the discrepancy between the predictions of the two methods below threshold.  As explained above, the stimulated emission {\it ignition} requires the simultaneous presence of at least one spontaneous photon in the lasing mode and the ``positive'' outcome for the stochastically--controlled {\it conversion} of the spontaneous into a stimulated photon (which later allows $E_s$, eqs.~(\ref{MSS},\ref{sSS}), to take over) in the time interval $\tau$.  As explained in~\cite{Puccioni2015} additivity for the probability distributions requires small arguments, thus low probability for conversion in the single time step $\tau$.  It is therefore plausible that the average contribution of the stimulated photons be small below threshold, and that it should remain that way until close to the transition.  It is indeed the essence of a (``phase'') transition to abruptly allow the buildup of the conditions which realize the change in state.  This explains why the SLS predicts a smaller photon average (but with much larger fluctuations!, cf. Fig.~\ref{numspikes}) than the REs, which ignore the conversion efficiency. 

The degree of quantitative reliability of the SLS predictions is a point which requires additional investigation.  The previous considerations based on the stimulated emission rate, $R_{st}$, well-known from the literature, are certainly physically meaningful and are capable of reproducing experimental observations (cf. Figs.~\ref{spikes} and~\ref{numspikes}a) which cannot be predicted by the REs (Fig.~\ref{numspikes}b).  In addition, the sharp change in the proportion of (average) stimulated vs. spontaneous photons at $C \approx 0.98$ matches the observations which come from the stability and eigenvector analysis (section~\ref{topology}).  The aspect which requires a closer look is whether the quantitative ratios of average stimulated and spontaneous photons predicted by the SLS, as a function of pump, are realistic or need improvement in the estimates of the stimulated emission rate.  In the RE approach, however, these considerations are not taken into account, thus leading to the mixing of the two kinds of photons, since the stimulated ones are supposed to be converted with probability 1 in all cases.  

The failure of REs to correctly describe the dynamics can be ascribed precisely to the assignment of perfect and instantaneous conversion of all spontaneous photons coupled into the on-axis mode into stimulated photons.  This hypothesis is, one the one hand, physically unrealistic since it contradicts the Fermi golden rule~\cite{CohenTannoudji} on timescales which are physically relevant, and, on the other hand, removes the physical mechanism for the occasional (i.e., stochastically--induced) accumulation of excess population inversion, responsible for the buildup of spontaneous photon bursts.  This answers the question asked in section~\ref{contradictions} concerning the origin of the spontaneous photon bursts:  the partial conversion and the stochastic delays lead to accumulation of energy into the carriers, with the consequent release of photon bursts.  

At the same time, the correctness of the global physical description of the REs (valid also for the SLS) is reflected by the possibility of correctly describing self-pulsing~\cite{Wang2017}, with the reintroduction, ``by hand'', of the mechanism responsible for the accumulation of population inversion~\cite{Wang2017}. 

An interesting consequence of the apparently minor, but conceptually major, modification in the description of stimulated emission introduced by the SLS concerns the estimates of the $\beta$ parameter from the shape of the characteristic laser curves.  One commonly accepted way is to estimate the jump in average number of emitted photons between the lower and the upper branch.  This way, $\beta^{-1} = \langle m_{upper} - m_{lower} \rangle$ ($m$ representing the total photon number in the lasing mode), obtained by extrapolating the two branches and estimating the jump at the threshold value.  This estimate is, however, based on the RE approach, which assigns probability 1 of conversion to all spontaneous photons coupled into the lasing mode, irrespective of the pump value.  If deviations from this representation prove to be sizeable, then the $\beta$--estimates would also be affected.

\section{Spiking in Class A lasers}\label{Aspikes}

Intuitively, one may attribute the existence of the discussed photon spikes uniquely to the interplay of the slow population inversion with the fast, but necessarily delayed, action of the electromagnetic field.  Recent theoretical work has shown that this is actually not the case and that large spikes are to be expected in Class A lasers as well~\cite{Vallet2019}.  The relevance of this finding is double, since on the one hand it deepens our understanding of the lasing transition, on the other hand it confirms that stochastic approaches (supported by analytical considerations, as in this manuscript) are best suited for advancing our understanding.  Fully stochastic simulations, pioneered about ten years ago~\cite{RoyChoudhury2009,RoyChoudhury2010} for lasers close\footnote{The statement is based on some indirect information obtained from the papers and on the comparison with the results of~\cite{Vallet2019}, given that the publications do not specify all crucial parameters.} to Class A and based on Monte-Carlo techniques, show the appearence of photon bursts below the continuous emitting regime (thus below threshold).  More sophisticated Markov simulations~\cite{Vallet2019} based on~\cite{Chusseau2002,Chusseau2014}, which include in the model the electronic bands in a semiconductor in great details and compute the carrier distributions, confirm the existence of such pulses, with very large photon spikes.  

In~\cite{Vallet2019} the cavity photon lifetime is chosen to be ten times larger than the carrier's relaxation, thus placing the system in a truly Class A regime.  The fast variable which deviates from equilibrium is now the population which can quickly grow away from its steady state if the stimulated emission process does not take hold.  While not accountable by a differential mathematical model using only the field variable~\cite{Narducci1988}, the intrinsic (and stochastic) delay with which the stimulated emission is {\it ignited} -- intrinsically included in the consistent stochastic representation -- allows for the accumulation of an excess energy in the laser, which is then released (``slowly'', due to the slower photon dynamics) as a spontaneous photon pulse.

Several of the conclusions reached in~\cite{Vallet2019} are confirmed by the analysis of~\cite{Takemura2019,Takemura2019B} based on a Fokker-Planck representation and on its numerical predictions.  Interestingly, a phase space analysis, similar to the one we have presented in this manuscript, is carried out~\cite{Takemura2019B} and shows the changes in the phase space properties which take place when passing from a strongly class A to a strongly class B device.  Of particular interest is the intermediate condition, where the relaxation rates of cavity and population are the same.  The conclusions of these investigations, based on complementary approaches, support one another and help shedding new light into 
the physics of the laser threshold.

\section{Conclusions}

Dynamical measurements of the photon emission in a mesoscale ($\beta = 10^{-4}$) laser at the onset of lasing show the occurrence of large, spontaneous photon bursts, which cannot be explained by Rate Equations (with Langevin noise).  Predictions based on the Stochastic Laser Simulator are instead capable of accounting for the existence of these bursts, based on the physical constraint that stimulated emission -- although characterized by a rate much faster than spontaneous relaxations -- is not an instantaneous process.  The stochastically--induced delay in the conversion of spontaneous into stimulated photons, with the consequent accumulation of excess population inversion, is responsible for the occasionally observed bursts of emitted photons, which correspond to stochastic occurrences where the conversion has failed for a sufficiently long time to generate the equivalent of a spontaneous gain-switch.

These considerations, which we have carried out on a low-$\beta$ device, are more general.  Stochastic simulations indeed predict pulsing before the lasing threshold for $10^{-4} < \beta < 0.1$ (cf.~\cite{Puccioni2015}, Fig. 3, and Supplementary Information in~\cite{Wang2015}, Fig. 5).  Experimental and numerical observations for the same class of low-$\beta$ lasers have already shown a link between the degree of correlation ($g^{(2)}(0)$) and the coherence (highlighted by $g^{(2)}(\tau)$) in the pulsing regime~\cite{Wang2015}, while future work will be necessary to establish a general scaling as a function of cavity volume.

Furthermore, we have examined the interpretation of the average photon number in the laser, as predicted by the Laser Rate Equations, exploring its limitations but also its strengths.  Their description is in fact closely linked to the Stochastic Laser Simulator since the latter stems from the same semi-classical representation of the radiation--matter interaction.  Thus, while the dynamical characteristics of Rate-Equations-based predictions are incorrect, nonetheless, the topological features are well-captured by the differential approach and fully explain the stochastically predicted, and experimentally observed, dynamics.  This paper thus fully shows the need for including the role of the finite stimulated emission rate in the transition from the incoherent to the coherent lasing regime.

As a final remark, it is interesting to notice that the physical conditions which lead to the appearence of the spontaneous spiking regime can be exploited for practical purposes.  Modulating the laser across the pump range where spikes exist, it is possible to generate trains of pulses which are quite regular at a frequency limited by the intrinsic system's response~\cite{Wang2019}.  In the presence of low incoherent feedback, on the other hand, it is possible to obtain an enhancement of the spectral components due to external cavity accompanying the spikes, together with the usual increase in output power originating from feedback~\cite{Wang2019b}. 

\section*{Acknowledgments}

T.W. is grateful to the R\'egion PACA (COSOUMISE project) and NSFC (61804036) for funding.  BBright has supported this project.  G.L.L. acknowledges discussions with A. Carmele, L. Chusseau, M. Gattobigio and L. Gil.

\end{document}